# Light chain systemic amyloidosis manifested as liver failure complicated with fatal spontaneous splenic rupture: A case report

Li Duo[1], Zhao Yingren[1], Chen Hongmei[1*]

[1]Xi'an Jiao Tong Univsity, Affiliated Hosp1 ,shaanxi,Peoples R China

**[Abstract]**

For a patient with manifestations of nausea, abdominal distension, spontaneous splenic rupture, obvious liver enlargement, low red blood cells and platelets, yellow sclera, and spider angioma, Congo red staining of liver and spleen tissues indicated amyloidosis. After secondary factors were excluded, the patient was finally diagnosed as chronic liver failure, light chain amyloidosis, spontaneous bacterial peritonitis after cystic resection and splenectomy. This case suggests that for patients with chronic liver failure accompanied by spontaneous splenic rupture and hepatomegaly for unknown reasons, the possibility of amyloidosis should be considered after excluding other factors, such as viral liver disease, autoimmune disease, alcoholic liver disease, genetic metabolic liver disease and liver tumor, and etc. Considering the low clinical incidence rate and poor prognosis, relevant diagnosis depends on the biopsy results, and many patients were not confirmed until autopsy after death. Therefore, once amyloidosis is suspected, it is necessary to have communications with relevant patients and their families on the risks for examination, treatment methods and prognosis as soon as possible.

**[Key words]**

Liver failure; Spontaneous splenic rupture; Light chain amyloidosis; Congo red staining

The pathological mechanism for amyloidosis is as follows: Amyloid deposits between the cells in various organs, which may lead to dysfunction of the cells, and eventually cause failure of the affected organs [1]. Amyloid is essentially a protein binding to mucopolysaccharides, which turns blue when exposed to iodine, just the same as starch. It is not tissue or organ specific, and may be found anywhere in the body. Therefore, relevant clinical manifestations are complicated, and easy to be misdiagnosed. There following is a report concerning a case of light chain systemic amyloidosis manifested as liver failure complicated with fatal spontaneous splenic rupture, who was treated in our department.

**[General information]**

Medical history

The patient, Last name: Shi, male, 50 years old, living in Tongchuan City, China, has been engaged in automobile repair for 30 years. He was admitted to our department on June 13, 2020 for chief complaint of "1 year of abdominal distension, accompanied with yellowing of eyes, which was aggravated with edema of lower limbs 20 days ago". One year ago, the patient developed abdominal distension, yellowing of eyes, fatigue, nausea without vomiting, abdominal pain and diarrhea, and body weight loss of about 15 kg. He was treated in a local hospital, Xijing Hospital, Fengxian Hospital and our hospital successively, and checked for liver function. The results were as follows: albumin 27.2g/L, TBil 44.9umol/L, ALT 14IU/L; AST:45IU/L. Gastroscopy found reflux esophagitis (Grade A), superficial gastritis with erosion. Ultrasonography showed no abnormality in thyroid function. Enhanced CT of upper abdomen found large liver with decreased density, liver cyst and several small retroperitoneal lymph nodes. Transient elastography of the liver showed a median liver hardness of 25.9 kPa and a median fat attenuation of 21. dm /m, for which the patient was diagnosed as "fatty liver" and given oral Methimazole, Ursofalk, Wuling Capsules, Diammonium Glycyrrhizinate Enteric-coated Capsules, and Glutathione Tablets, as well as a traditional Chinese medicine for twice (not specified), after which abdominal distension and yellowing of eyes were alleviated. Twenty days ago, his abdominal distension was aggravated for unknown reasons, and there was edema of lower limbs after activity, which could be relieved after rest. He felt that the lower limbs "had no nowhere to place", with poor sleep at night, and then, he was treated at the Second Affiliated Hospital of Xi 'an Jiaotong University. Relevant examination results were as follows: Lier function: albumin 33.5g/L, TBil: 91.20umol/L, ALT: 27IU/L; AST: 85IU/L, Hepatic angiography findings: The posterior hepatic segment of inferior vena cava was thin, and the intrahepatic vein was unclear, with enlarged liver and ascites. Chest CT findings: chronic inflammation in the upper lobe of right lung, and in the left lung; right pleural effusion. Bone biopsy: myelodysplastic syndrome with abnormal lymphocytes, for which the patient received supportive therapies, such as liver protection and yellowing alleviation, ammonia reduction, albumin infusion, plasma and diuresis. After this, the abdominal distension and edema of lower limbs were alleviated, but the sleep was still poor at night. Then he came to our hospital for further diagnosis and treatment. Since the onset of the disease, his mental conditions, as well as his appetite, were poor, his stool was dry and the urine was yellow and heavy. Also, his sleep was poor at night. Past history: The patient received cholecystectomy more than 1 year ago, and splenectomy for spontaneous rupture of the spleen about 5 months ago, with blood transfusion history. He denied the histories of hepatitis B, hepatitis C, tuberculosis, malaria, hypertension, heart disease, diabetes, cerebrovascular disease, and mental disease. He had 30 years of smoking history, average 2 cigarettes/day, but has given up smoking. Also, he had 25 years of drinking history, 2-3 times a week, about 150g/ time, but has given up drinking. Family history: His father died for renal failure, with no special epidemiological history.

Physical examination

Vital signs after admission: T: 37.2°C, P: 101 times/min, R: 20 times/min, BP :124/78mmHg, height 170cm, weight 62kg. With appearance of liver disease, autonomous position, conscious, slow in response, good in mental conditions, without yellow stain or rash on skin of the whole body, with ecchymosis at the puncture point on the forearm, 1 spider mole on the chest, and liver palms. The sclera was slightly yellow and the eyelids was pale, with enlarged lymph nodes palpable in the neck; Oral cavity: There were ulcers at the buccal mucosa and lower lip, and pigmentation on the right tongue bod. There was no thyroid enlargement, and no vascular murmur was heard. Chest: The chest was symmetrical without deformity, without obvious abnormality of respiratory movement, and the vocal fremitus was not enhanced or weakened. Double lung percussion, as well as double lung auscultation for breath, sounds clear. Dry and wet rales and pleural frictions were not heard. Heart: There was no eminence in the precardiac area, and there was cardiac apex fluctuation 0.5cm in the midline of the left fifth intercostal clavicle. No tremor were touched and pericardial friction sound. Percussion showed not cardiac enlargement. The heart rate was 101 times/min, with regular hear rhythm. No murmur was heard in each valve area. Abdomen: Abdominal distension, without abdominal varicose veins, with about 15cm surgical scar under the left rib. There were tenderness below the liver margin and under the xiphoid process, without rebound pain, and the liver is 8cm below the ribs, spleen was not touched, with abdominal dull percussion note, and positive shifting dullness. The auscultation showed weakened borborygmus. Nervous system: asterixis, ankle clonus (-). Initial diagnosis: chronic liver failure, damage for alcoholic liver? Hepatic encephalopathy? spontaneous bacterial peritonitis after splenectomy and cholecystectomy.

Auxiliary examination

The laboratory examination results after admission in our hospital were as follows: routine blood test: HGB92g/L, WBC 16.17×109/L, N% 74.4%, L%17.2%, PLT83×109/L. Hepatic function: ALB37.3g/L, AST60U/L, ALT30U/L, CHOL1.24umol/L, TBIL142.2umol/L. Blood coagulation: PT:28.60S, APTT62.40S, INR2.66, PTA:27% , FIB1.45g/L. Myocardial enzyme spectrum: ADH390U/L, HBDH378U/L, CK291U/L, CK-MB 27U/L, IMA 61.8U/ml. ECG: sinus rhythm, low voltage, ST-T segment changes; Renal function: cysC2.000mg/L Cr68umol/L, BUN 11.76mmol/L; ESR: 54mm/h; PCT: 0.818 ng/ml. No obvious abnormalities were observed in 10 indicators for tumor, 8 indicators for infection, routine feces and urine tests, plasma ammonia, G/GM, auto-immune liver antibody, ceruloplasmin, and ANCA. Hepatitis virus series of HBsAg, HBeAg, Anti-HBe, Anti-HBc (-), Anti-HBs(+); Hepatitis A, C and E antibodies: negative. EB-IgM (-), EB-IgG(+); EBV, CMV quantitative tests: negative. Among the immune indexes, there were eight for immunity and 3 for rheumatism: ASO 327IU/ mL, KAP light chain 7.58g/L, LAM light chain 4.32g/L, IgE1360IU/ml, IgG23.10g/L, IgA11.60g/L, C3:0.47g/L. Immunofixation electrophoresis: IgG, IgA, IgM, K, L: negative; Free light chain: KAP light chain 77.7g/L; LAM light chain: 80.30g/L; Lymphocyte subsets: total T cell percentage: 53.67%; NK cell percentage 38.51%; Helper T cell percentage: 21.26%; B cell percentage: 7.26%; Blood β 2 - ug 3388.7 MG/L; Urine β 2 - ug 50.0 MG/L; Serum protein electrophoresis: ALB: 51.50, α1-globulin 3.50, α2-globulin 3.50; β1-globulin 5.00, β2-globulin 12.10, γ-globulin 24.40; M protein 0.00; IgG typing: IgG18.90g/L; IgG1:16.90 g/L; IgG4:1.17 g/L; Imaging findings: MR plain scan of the head: no obvious abnormalities in brain parenchyma. Vertebral body MR: C3-7 and L5-S1 intervertebral disc degeneration and herniation, without obvious abnormalities in thoracic vertebra and thoracic pulp; Cardiac Ultrasonography: Left ventricular soothing function decreased; Enhanced CT for the upper abdomen: enlarged liver, sign of heterogeneous fatty liver, and slight hypodense lesion in the right anterior lobe of liver. The gallbladder and spleen were not clearly displayed. No obvious abnormal enhancement was observed in the pancreas, bilateral kidneys and bilateral adrenal glands (for changes in the liver and spleen, see Figure 1). Angiography for abdominal great vessels: No obvious abnormalities were found in the proper hepatic artery, and the left and right hepatic arteries, as well as in the superior mesenteric vein, residual splenic vein, main portal vein, and the left and right branches of portal vein. Localized stenosis was observed in the intrahepatic segment of inferior vena cava. Re-examination results of the spleen and liver biopsy specimens showed that the white pulp structure of spleen amyloidosis disappeared, with significant changes for congestion, local massive bleeding, and red stained vascular wall without structural deposits. Cargo red staining indicated amyloidosis. Immunohistochemical results of liver: CD34 staining showed local hepatic sinus endothelium vascularization, CK7 staining showed notable hyperplasia of bile canaliculus, HBsAg (-), special staining results: Masson, PSA, PAS-D, iron staining showed no abnormal local destruction of mesh fibers, Congo red (+). (See Fig. 2 and Fig. 3)

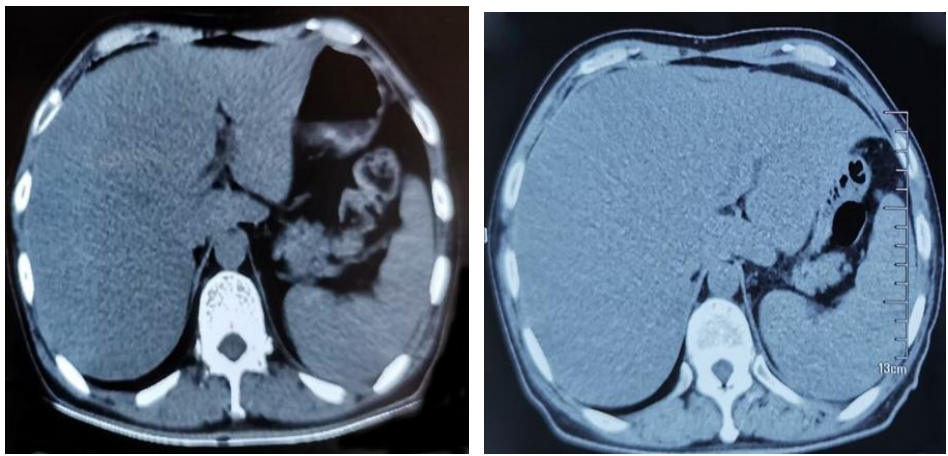

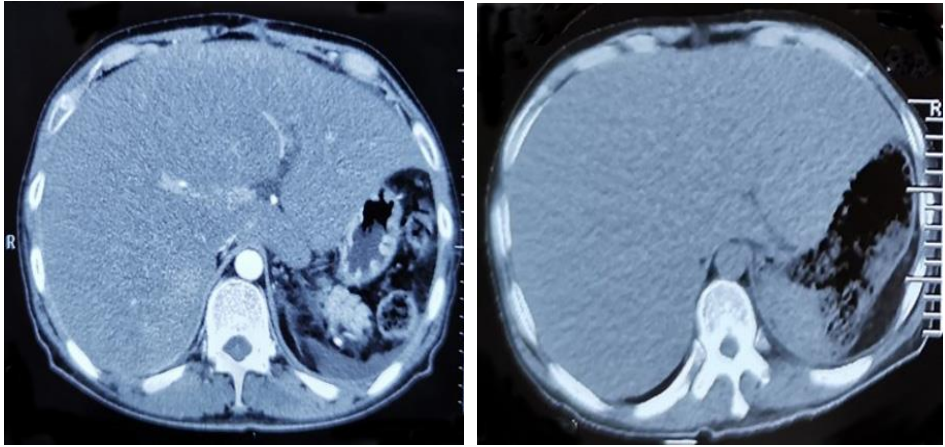

Fig. 1 Changes of liver and spleen

Note: The time changes of top left, top right, bottom left and bottom right are 2019-07-24, 2020-01-07, 2020-04-09 and 2020-06-12 respectively.

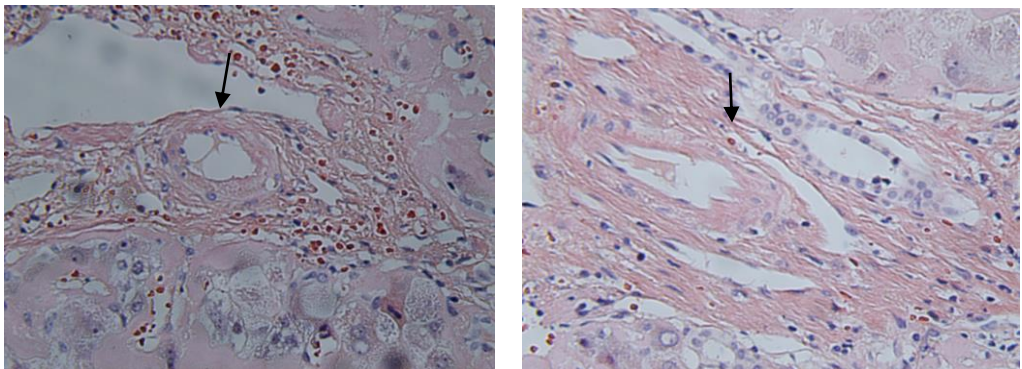

Fig. 2. Pathological change of the liver

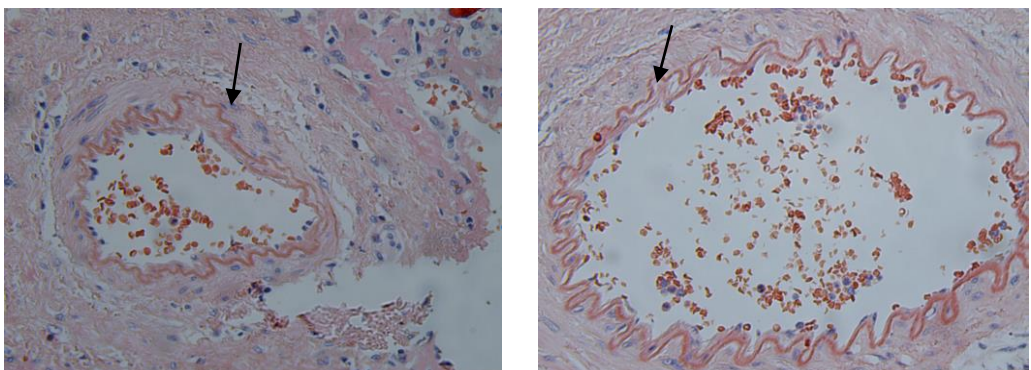

Fig. 3 Pathological change of the spleen

Summary

After admission to our hospital, the patient's manifestations in the central nervous system were as follows: He was slow in response, occasionally raving, but with asterixis and ankle clonus negative, plasma ammonia normal, for which hepatic encephalopathy was excluded, and nerve damage was highly suspected. Peripheral nervous system: Extreme fatigue of lower extremities, for which the patient felt that the lower limbs "had no nowhere to place", and his sleep was seriously affected, but the head MRI and lower extremities EMG showed no obvious abnormalities. C3-7 and L5-S1 intervertebral disc degeneration and herniation, without obvious abnormalities in thoracic vertebra and thoracic pulp, for which after consulting the Psychosocial Department, Neurology Department and Orthopedics Department were consulted respectively, the patient was diagnosed as "restless leg syndrome", with involvement of peripheral nerve very likely. Cardiovascular system: The patient developed left chest pain during hospitalization. Re-examination of ECG showed sinus rhythm, and generally, the ECG results were normal. There was no obvious abnormality in myocardial enzyme profile. Cardiac ultrasonography showed that the left ventricular soothing function was reduced. The chest pain could not exclude the possibility of myocardial amyloidosis. Blood system: Peripheral blood: progressive decrease of red blood cells and platelets; high KAP light chain and LAM light chain. Bone marrow: signs of hyperplasia, megaloblastoid changes of granulocytes and erythrocytes, increased proportion of reticular cells, increased peripheral leukocytes, visible abnormal lymphocytes, both cell lines reduced after splenectomy, which might be caused by liver damage, but in combination with bone marrow examination and related immune examination, myeloma could not be excluded. Kidney: Blood β2 microglobulin increased, with urine β2 microglobulin and 24h urine protein negative. eGFR: 140.31 ml/min / 1.73 m$^2$ (MDRD method). Enhanced CT showed no obvious abnormality. Angiography showed no obvious abnormality in renal morphology, indicating that the kidney was not involved. Liver: Progressive decline of liver function. CT showed hepatomegaly, diffuse low-density area, and no obvious enhancement. The CT value was lower than that of spleen, and similar to that of diffuse invasive hepatic fat infiltration. The spleen ruptured spontaneously on 2020-1-24. Re-examination of spleen and liver biopsy specimens showed Congo red (+), with amyloidosis of both liver and spleen. In combination with laboratory examination of free light chain: KAP light chain 77.7g/L; LAM light chain: 80.30g/L, the patient was diagnosed as light chain amyloidosis at last.

Problems in the systemic review on changes of the patient's conditions in the past year: First, during splenomegaly, ascites, and spider nevus as typical manifestations of decompensation in cirrhosis, there appeared giant liver as clinical manifestation, which was significantly inconsistent with liver cirrhosis. Second, after splenectomy, routine blood test showed that HGB and PLT were still decreasing. Third, there was no exact cause or obvious trigger for rapid decline of liver function. Fourth, the true cause for spontaneous splenic rupture. Furthermore, the incidence rate of amyloidosis was low, relevant experience was short, and it could not be identified by imaging alone. Therefore, as a rare disease, amyloidosis should be considered after excluding viral hepatitis, alcoholic liver disease, autoimmune liver disease, metabolic liver disease, biliary tract disease and liver damage caused by parasites.

**Discussion**

At present, the etiology or pathogenesis for amyloidosis is still unclear. The study of Shientag Lisa J et al. showed that in birds, R52L mutation in the SSA gene might be a genetic factor of hepatic amyloidosis[2]. And all diseases depend on the proportion and interaction of genetics and environment. There is still no report concerning studies on genetic polymorphism in humans for the low incidence rate and poor prognosis. Mahesh Lakmal Gunathilaka et al reported a case of chronic liver damage possibly caused by gasoline[3]. And in terms of environmental factors, the case had been engaged in the automobile repair for a long time. People have pay attention to the association of gasoline with human diseases, including liver disease, as well as the mutation of related genes. For amyloidosis, the early clinical manifestations were atypical, and relevant symptoms were various, and so, we need to inquire the medical history repeatedly, and perform check-ups carefully. In order to reduce the rate of misdiagnosis, it is necessary to be familiar with the diagnostic process of rare diseases based on understanding of common diseases and their etiology. For the treatment of light chain amyloidosis, bortezomib in combination with dexamethasone or cyclophosphamide has been considered, but the prognosis remains poor.

This case is a middle-aged male patient treated in several hospitals for recurrent nausea, abdominal distension, yellowing of the sclera, edema of lower limbs, and poor sleep. During the course of disease, spontaneous splenic rupture occurred for unknown reasons, for which we exerted efforts to found the reasons. The hemogram showed a progressive decline of red blood cells and platelets. The main manifestations for liver function were continuously increased ceramic oxalacetic transaminase and total bilirubin, and decreased albumin and total cholesterol, for blood coagulation function, progressive aggravation. The patient denied histories of hepatitis B, hepatitis C, tuberculosis, malaria, and contact with epidemic water. B-ultrasonography of the liver indicated that the liver was being enlarged. Transient fibrous imaging of the liver showed a median liver hardness of 25.9 kPa and a median fat attenuation of 21. dm /m. The reexamination of hepatitis virus in our hospital showed no abnormality in autoimmune liver antibody and ceruloplasmin. Pathological re-examination of spleen and liver pathology confirmed the diagnosis of light chain amyloidosis with chronic liver failure. The patient is still under follow-up.

**\*Corresponding author. Address:** Xi'an Jiao Tong Univsity, Affiliated Hosp1,Dept infect Dis, Coll Med,277 Yanta West Rd, Xian 710061,shaanxi,Peoples R China
**E-mail address:** chenhongmei1107@126.com

## Conflict of interest

The authors who have taken part in this study declared that they do not have anything to disclose regarding funding or conflict of interest with respect to this manuscript.

## Acknowledgements

The authors thank the patient and the patient's relatives for their participation. We would like to acknowledge the expert technical assistance of Sui Yanxia and Sun Xingwang who are from the Department of Pathology and Imaging Department of the First Affiliated Hospital of Xi'an Jiaotong University respectively.